\documentclass{osa-article}
\journal{osajournal}

\usepackage[utf8]{inputenc}
\usepackage{siunitx}             
\usepackage{graphicx}            
\usepackage{xcolor}   
\usepackage{amsmath}
\usepackage{amssymb}
\usepackage{mathtools}
\usepackage{braket}
\usepackage[version=3]{mhchem}
\usepackage{float}
\usepackage{hyperref} 
\usepackage{siunitx}
\hypersetup{colorlinks=true, citecolor=blue, colorlinks, linkcolor=blue, urlcolor=blue}

\usepackage{tikz,xcolor,hyperref}

\definecolor{lime}{HTML}{A6CE39}
\DeclareRobustCommand{\orcidicon}{
	\begin{tikzpicture}
	\draw[lime, fill=lime] (0,0) 
	circle [radius=0.16] 
	node[white] {{\fontfamily{qag}\selectfont \tiny ID}};
	\draw[white, fill=white] (-0.0625,0.095) 
	circle [radius=0.007];
	\end{tikzpicture}
	\hspace{-2mm}
}

\foreach \x in {A, ..., Z}{\expandafter\xdef\csname orcid\x\endcsname{\noexpand\href{https://orcid.org/\csname orcidauthor\x\endcsname}
			{\noexpand\orcidicon}}
}

\begin{document}

\title{A tellurite glass optical microbubble resonator}

\author{J. Yu,\authormark{1,3}\orcidB{} J. Zhang,\authormark{1}\orcidD{} C. Wang,\authormark{1} A. Li,\authormark{1,3}\orcidE{} M. Zhang,\authormark{1} S. Wang,\authormark{1}\orcidF{} P. Wang,\authormark{1,2,*}\orcidC{} J. M. Ward\authormark{3,4} and S. {Nic Chormaic}\authormark{3,+}\orcidA{}}
\address{
\authormark{1}Key Laboratory of In-Fiber Integrated Optics of Ministry of Education, College of Science, Harbin Engineering University, Harbin 150001, China\\
\authormark{2}Key Laboratory of Optoelectronic Devices
and Systems of Ministry of Education and Guangdong Province College of
Optoelectronic Engineering, Shenzhen University, Shenzhen 518060, China\\
\authormark{3}Light-Matter Interactions for Quantum Technologies Unit, Okinawa Institute of Science and Technology Graduate University, Onna, Okinawa 904-0495, Japan\\
\authormark{4}Physics Dept., University College Cork, Cork, Ireland\\

}

\email{\authormark{*}pengfei.wang@tudublin.ie}
\email{\authormark{+}sile.nicchormaic@oist.jp}

\begin{abstract}
We present a method for making microbubble whispering gallery resonators (WGRs) from tellurite, which is a soft glass, using a CO$_2$ laser. The customized fabrication process permits us to process glasses with low melting points into microbubbles with loaded quality factors as high as $2.3 \times 10^6$. The advantage of soft glasses is that they provide a wide range of refractive index, thermo-optical and optomechanical properties. The temperature and air pressure dependent optical characteristics of both passive and active tellurite microbubbles are investigated. For passive tellurite microbubbles, the measured temperature and air pressure sensitivities are 4.9~GHz/K and 7.1~GHz/bar, respectively. The large thermal tuning rate is due to the large thermal expansion coefficient of $1.9 \times 10^{-5}$ K$^{-1}$ of the tellurite microbubble. In the active Yb$^{3+}$-Er$^{3+}$ co-doped tellurite microbubbles, C-band single-mode lasing with a threshold of 1.66~mW is observed with a 980 nm pump and a maximum wavelength tuning range of 1.53~nm is obtained. The sensitivity of the laser output frequency to pressure changes is 6.5~GHz/bar. The microbubbles fabricated by this novel method have a low eccentricity and uniform wall thickness, as determined from electron microscope images and the optical spectra. The compound glass microbubbles described herein have potential for a wide range of applications, including sensing, as tunable microcavity lasers, and for integrated photonics.
\end{abstract}

\bibliography{ref}
\section{Introduction}
\label{sec:Introduction}
In the past few decades, the level of research activity on whispering gallery mode (WGM) resonators, has increased rapidly\cite{Cai:00, Rezac:01, White:05, Ooka_2015, Bianucci_2016, ShoComb2018, FRUSTACI201966}. Concurrently, the number of WGM geometries has also increased. These resonators, or microcavities, have small sizes, high uniformity and smooth surfaces which combine to deliver an extremely high quality (\textit{Q-}) factor and small mode volume, thereby having potential in many applications, such as high-sensitivity sensing, nonlinear optics, optomechanics, active photonics devices, cavity quantum electrodynamics, and nanoparticle control\cite{Kippenberg555, Bahl201, PhysRevA.90.053822, Foreman:15, doi:10.1063/1.4922637, Yang:16, Yang:116, Yang:17, Ward:18, doi:10.1021/acsnano.9b04702}. One of the resonator geometries that has attracted much attention is the microbubble, in which the WGMs propagate in the wall of a thin spherical shell typically made of glass. First reported in 2010 \cite{Sumetsky:s}, a silica microbubble can be fabricated by heating and expanding a silica capillary using a CO$_2$ laser and air pressure \cite{Watkins:s}. The wall thickness of the bubble can be close to the wavelength of light propagating in the WGMs, resulting in evanescent fields on the inner and outer walls; therefore, the modes are extremely sensitive to changes in refractive index. The thin walls also give a large sensitivity to changes in pressure \cite{Yang:16}.

Microbubbles are predominately made from silica glass because it is relatively easy to manipulate when heat-softened. Other soft glasses have many advantages over silica, but their low melting points makes them more difficult to control and cast into the desired shape. In previous work, lead silicate microbubbles with single and double stems were fabricated. However, the shapes were not spherical, the Q factors were limited to about $10^5$, and the fabrication process was very difficult to control with a low success rate \cite{doi:10.1063/1.4908054}. In this work, we report on the development of a microbubble fabrication method for soft glasses to yield doped and undoped tellurite glass microbubble whispering gallery resonators (WGR). Tellurite glass has many properties which may lend themselves to the functionality of the WGR. For example, its thermal expansion coefficient \cite{INOUE2003133} is larger than that for silica \cite{White_1973}, thereby increasing the sensitivity of temperature sensors. This is also true for the nonlinear and optomechanical coefficients of tellurite glass. One of the benefits of laser emission in a microbubble is the high degree of wavelength tunability and the prospect of integrating a microlaser into a hollow WGR, but, at present, the common fabrication method is to cover the surface of the microcavity with a gain material, such as sol-gel or some other rare-earth ion doped compound glass \cite{Ward2016, Yang:17} --- a review of several techniques is contained in \cite{Righini_2016}. An alternative is to inject a gain liquid (such as an organic gain dye) into the microbubble for laser emission \cite{doi:10.1063/1.3629814}. In this case, a large loss will be induced in the microbubble resonator since the introduction of a gain material, which is not intrinsic to the capillary, results in a higher lasing threshold. Additionally, the material of choice for the capillary has been predominantly silica, which is limited in its rare-earth ion doping concentration and low phonon energy, so it can be challenging to know the exact concentration of the gain medium after preparing the resonator. Finally, the wavelength range of laser emissions from silica is largely limited to visible and near-infrared light. 

The aforementioned drawbacks with existing techniques can be largely overcome by using compound tellurite glass to prepare the microbubble resonator, details of which are contained herein. A polished tellurite glass tube was initially formed into a microcapillary by tapering using a large fiber drawing tower. Then the capillary was further drawn down to its final diameter by heating and stretching it in a custom-made pulling rig consisting of a small ceramic heater. Next, a CO$_2$ laser was used to form the tellurite glass microbubble using a unique method. Finally, the temperature and air pressure dependent optical characteristics of the WGMs were investigated for both passive and active versions of the microbubble.

\section{Fabrication of the tellurite microbubble}
\label{sec:Setup}

As a first step, both passive and active tellurite glass rods, with composition 75TeO$_2$-5ZnO$_5$-15Na$_2$CO$_3$-5Bi$_2$O$_3$ and 75TeO$_2$-5ZnO$_5$-15Na$_2$CO$_3$-4.25Bi$_2$O$_3$-0.5Yb$_2$O$_3$-0.25Er$_2$O$_3$, respectively, were fabricated using the melt-quenching method \cite{TANABE2002815}. 50 g of high-purity chemical material [TeO$_2$ (99.99\%), ZnO$_2$ (99.99\%), Na$_2$CO$_3$ (99.99\%), Bi$_2$O$_3$ (99.99\%), Yb$_2$O$_3$ (99.99\%), Er$_2$O$_3$ (99.99\%)] was placed in an agate mortar and stirred for 10 minutes, stored in an alumina crucible, and heated in a closed furnace at $900^\circ$C for 60 minutes. Then the melt was poured quickly into a tube furnace, which had been heated to $300^\circ$C for 3 hours. The rotation speed of the tube furnace was set to 30~rev/min for one minute. A glass tube was formed in the tube furnace and annealed at $310^\circ$C for 4 hours to remove any remaining internal stress. Next, the glass tube was removed and polished using low-mesh to high-mesh sandpaper until the ratio of the inner and outer diameters was between 0.6-0.7. The prepared polished glass tube was mounted in the nitrogen-filled chamber of the fiber drawing tower. The temperature of the furnace was increased to 345$^{\circ}$C and tellurite glass capillaries with outer diameters of 300~$\mu$m were obtained by controlling the speed and the tractive force of the fiber drawing tower.
\begin{figure} [ht]
\centering
      \includegraphics[width=0.7\textwidth]{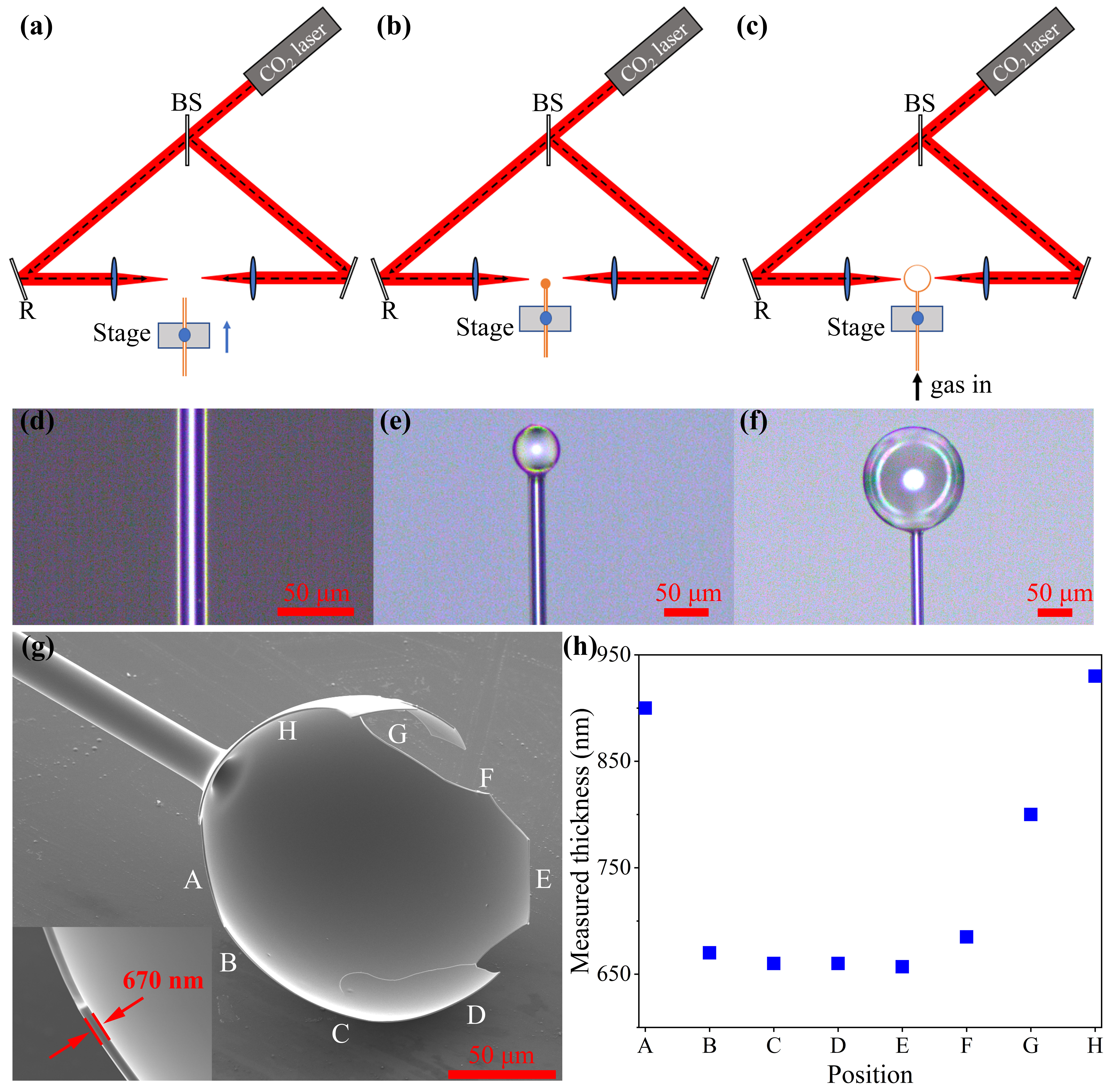}
  \caption{(a-c) Three step fabrication processes for the tellurite glass microbubble. BS: beam splitter, R: mirror. (d-f) Microscope images of the tellurite glass capillary, microsphere, and microbubble in steps (a-c). (g) SEM image of a broken tellurite glass microbubble, A-H represent the different measurement positions in the microbubble, with the wall thickness at position B being 670~nm. (h) The measured thickness at the different positions in (g). 
  \label{fig:fabrication}}
\end{figure} 

To make microbubbles, several steps are needed as illustrated in Figure \ref{fig:fabrication}(a)-(f). First, we need to further decrease the diameter of the tellurite capillaries to about 15~$\mu$m using a ceramic heater, see Figure \ref{fig:fabrication}(d). In order to get a tellurite microbubble, a CO$_2$ laser is needed to process the tellurite capillaries even further. The laser beam was divided into two parts, which were then focused at the same point, as shown in the schematic of Figure \ref{fig:fabrication}(a). A solid microsphere with a diameter of 35~$\mu$m was first formed on the tip of the tellurite glass capillary. This was done by affixing the capillary to a one-dimensional translation stage and then slowly moving it into the center of the two laser beams until the microsphere forms. At this point, a gas valve was opened and the capillary was pressurized to 3.5 bar, while, at the same time, the power of the CO$_2$ laser was increased. The tellurite microsphere was expanded into a microbubble with a diameter of about 150~$\mu$m, as shown in Figure \ref{fig:fabrication}(f). This is a self-terminating process;  once the CO$_2$ laser power and gas pressure are fixed, the bubble expansion stops at the point where the heat loss from the wall exceeds the absorbed heat from the laser.  This method is simpler and reduces cavity loss when compared with the method of directly making a microbubble from the capillary \cite{doi:10.1063/1.4908054}. During the glass capillary fabrication process, some tiny air bubbles are formed when the molten glass is rotated in the tube furnace. A rough surface is also created by uneven polishing. If a bubble is made directly from the capillary by simply softening the glass, the air bubbles and rough surface may be preserved. However, if the microcapillary is completely remelted into a microsphere, the bubbles inside  disappear and the surface is very smooth.

In order to characterize the wall thickness of the microbubble, it was broken and imaged with a scanning electron microscope (SEM). The measured thickness was about 670~nm at position B, as shown in Figure \ref{fig:fabrication}(g). We  measured multiple points to determine the uniformity of the microbubble and the results are shown in \ref{fig:fabrication}(h). Except for the thickness near the stem, which was around 900 nm, most other points on the microbubble wall were around 670 nm. The thickness of the wall was calculated to be 640~nm by assuming conservation of volume from the tellurite glass in the microsphere to the resulting microbubble and is consistent with the measured value \cite{Jiang:20}. The thickness was calculated from 
\begin{equation}
\frac{4}{3}\pi(\frac{c_1}{2})^3=\frac{4}{3}\pi(\frac{c_2}{2})^3-\frac{4}{3}\pi(\frac{c_2-a_1}{2})^3,
\end{equation}
where \textit{c$_1$} and \textit{c$_2$} are the diameters of the microsphere and microbubble, respectively, and \textit{a$_1$} is the thickness of the microbubble wall. The above formula is different from the mass conservation of the cross-sectional area described in the  literature \cite{Henze:11,Cosci:15}. Since the tellurite capillary tends to collapse to some degree when it is tapered with the ceramic heater, its accurate inner diameter could not be obtained.  The microbubble shown in Figure \ref{fig:fabrication}(g) differs from other conventional silica microbubbles reported in the literature \cite{Yang:16}, not only in the fabrication method but also the final geometry. The bubble is highly uniform both in shape and wall thickness. The fact that the microbubble is blown out from a microsphere means that the resulting bubble  also has a low degree of eccentricity. Microbubbles from other low-melting compound glasses, such as fluoride and chalcogenide glasses, could be prepared using this method.

\section{Experimental setup}
\begin{figure} [ht]
\centering
      \includegraphics[width=0.9\textwidth]{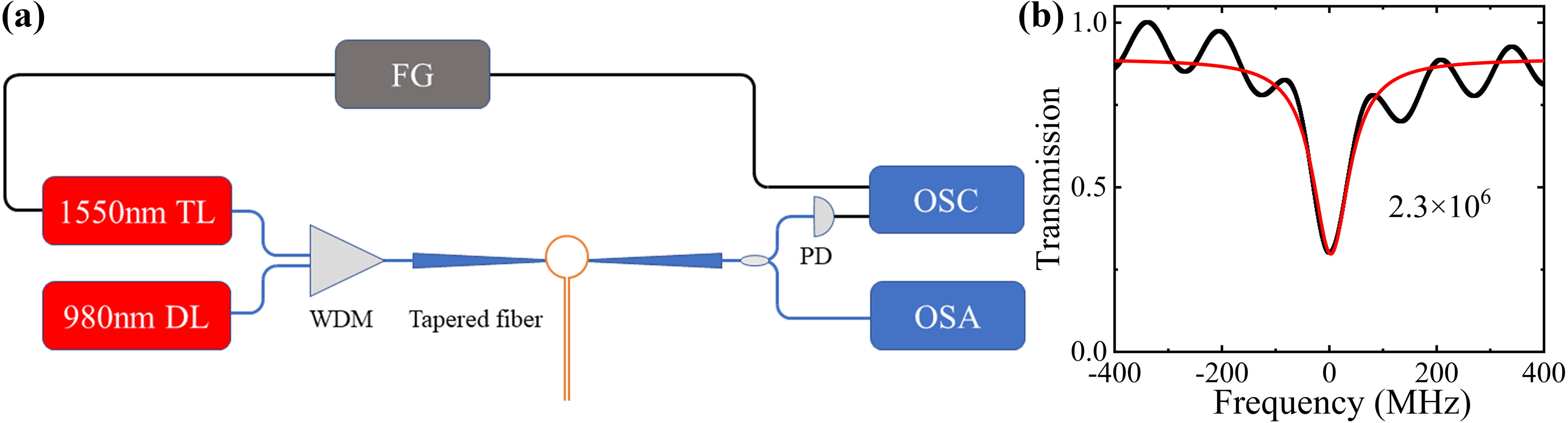}
  \caption{(a) Experimental setup for the tellurite glass microbubble. The blue line represents the optical path and the black lines are the electrical connections. TL: tunable laser; DL: diode laser; FG: function generator; WDM: wavelength division multiplexer; PD: photodetector; OSC: oscilloscope; OSA: optical spectrum analyzer. (b) The observed WGM resonance spectrum of the tellurite glass microbubble (diameter $\sim$ 130 $\mu$m, wall thickness $\sim$ 800 nm) with a Lorentzian fit (red line), corresponding to a loaded $Q$-factor of $2.3\times10^6$.}
  \label{fig:ExptRes}
\end{figure}

The experimental setup is schematically illustrated in Figure \ref{fig:ExptRes}(a). We used two lasers in a pump/probe arrangement. A tunable laser (TLB-6700, Newport) with a center wavelength of 1550~nm was used to probe the WGM resonances of the tellurite glass microbubbles. The laser frequency was scanned over 36~GHz at a rate of 10~Hz. A pump diode laser with a wavelength of 980~nm (BL976-SAG300, Thorlabs) and a linewidth of 1~nm was used to both control the temperature inside the passive tellurite glass microbubble and act as the pump for the Yb$^{3+}$-Er$^{3+}$ doped tellurite glass microbubble. A fiber with single-mode transmission at wavelengths of 980 nm and 1550 nm was selected (1060XP, Thorlabs) to make the tapered coupling fiber with a final diameter of around 1~$\mu$m. The output lasing was observed using an optical spectrum analyzer (MS9740A, Anritsu).

\section{Passive tellurite glass microbubbles}
\begin{figure} [ht]
\centering
      \includegraphics[width=0.9\textwidth]{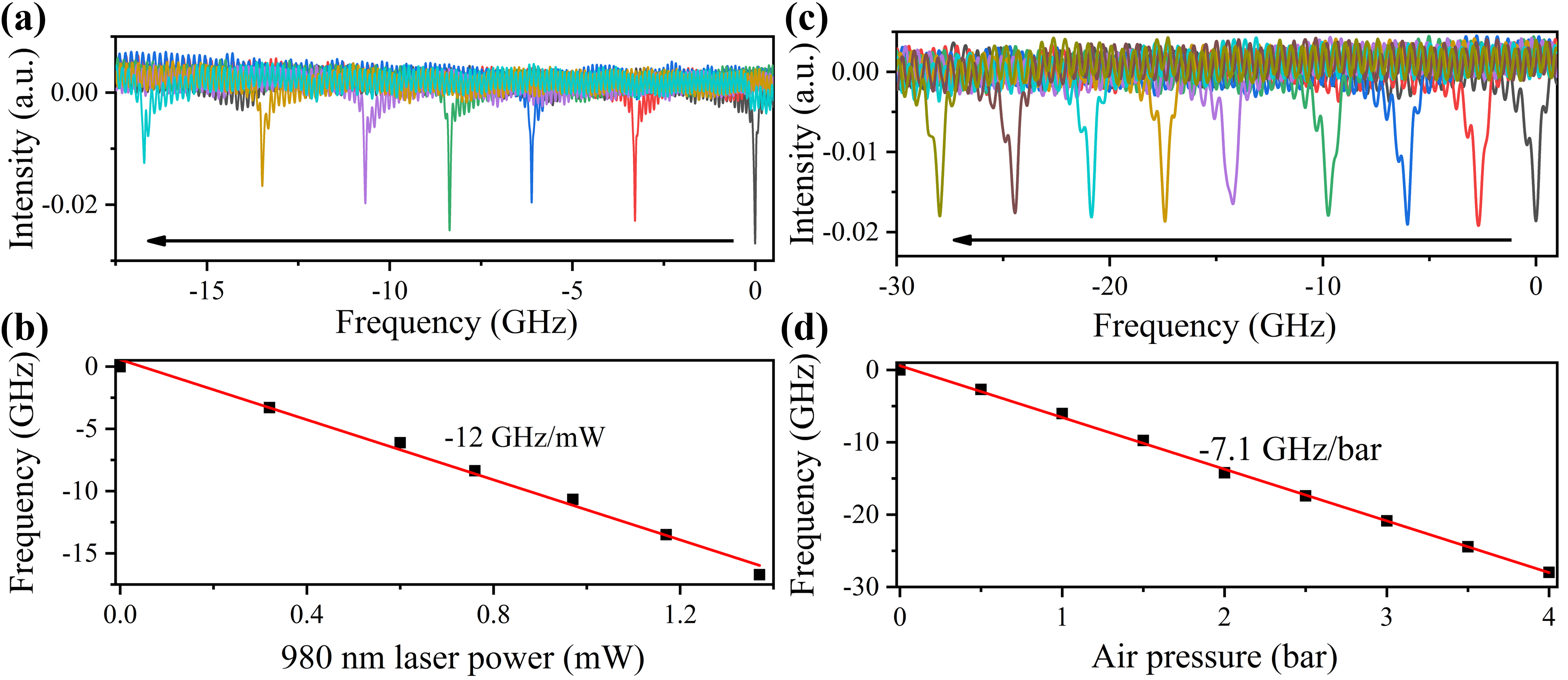}
  \caption{Resonance shift of the passive microbubble with varying (a) 980~nm laser power or (c) air pressure. The black arrow shows the direction of the resonance shift.  Resonance frequency shift as a function of (b) 980 nm laser power and (d) air pressure.  The red lines are linear fits to the experimental data. 
  \label{fig:shifts}}
\end{figure}

As a first step, the \textit{Q}-factor of a passive tellurite glass microbubble was measured by scanning the frequency of the tunable laser. A typical transmission spectrum as a function of laser frequency is shown in Figure \ref{fig:ExptRes}(b) and the fitted \textit{Q} is $2.3\times10^6$, which is close to the value of passive tellurite glass microspheres presented  previously \cite{8928617}. Next, we used the 980~nm diode laser to control the temperature inside the microbubble while the 1550~nm WGM resonance frequency was recorded. The results are shown in Figure \ref{fig:shifts}(a). As we increased the pump laser output power from 0 to 1.37~mW, the temperature of the glass increased due to absorption, resulting in a red shift of the resonance frequency, \textit{f}, by 16.7 {GHz}. A linear fit yields a power sensitivity of $-12$~GHz/mW. The total shift, $\Delta{f}$, of the frequency is given by

\begin{equation}
\Delta\textit{f}=\textit{f}(\frac{1}{n}\frac{\Delta\textit{n}}{\Delta\textit{T}}+\frac{1}{d}\frac{\Delta\textit{d}}{\Delta\textit{T}})\Delta\textit{T},
\end{equation}
where \textit{n} is the refractive index and \textit{d} is the diameter of the tellurite microbubble. The thermal expansion coefficient for tellurite glass is $1.9\times10^{-5}$ K$^{-1}$ \cite{INOUE2003133}, which is 38 times larger than the corresponding value for silica glass of $0.51\times10^{-6}$~K$^{-1}$ \cite{White_1973}. The thermo-optic coefficient $\Delta$\textit{n}/$\Delta$\textit{T} is $1.08\times10^{-5} $~K$^{-1}$ \cite{Kassab_2007}, and the frequency shift as a function of temperature $\Delta$\textit{f}/$\Delta$\textit{T} was calculated to be 4.9~GHz/K, which is about 4 times lager than for a silica microsphere of 1.28~GHz/K at room temperature \cite{Ma_2010}.

According to Equation (2), the frequency shift can also be affected by air pressure, which changes the diameter of the microbubble and the refractive index by stress, see Figure \ref{fig:shifts}(c). The resonance red shifted by 27.9~GHz as the air pressure inside the bubble was increased from 0 to 4 bar, yielding a pressure sensitivity of $-7.1$~GHz/bar. For a silica microbubble with a diameter of 141~$\mu$m and a thickness of 1.3~$\mu$m, the pressure sensitivity of the resonance is $-8.2$~GHz/bar \cite{Yang:s}, 
which is close to the value we have obtained for the tellurite glass microbubble with a diameter of 130~$\mu$m and a thickness around 800~nm herein.
\begin{figure} [ht]
\centering
      \includegraphics[width=0.5\textwidth]{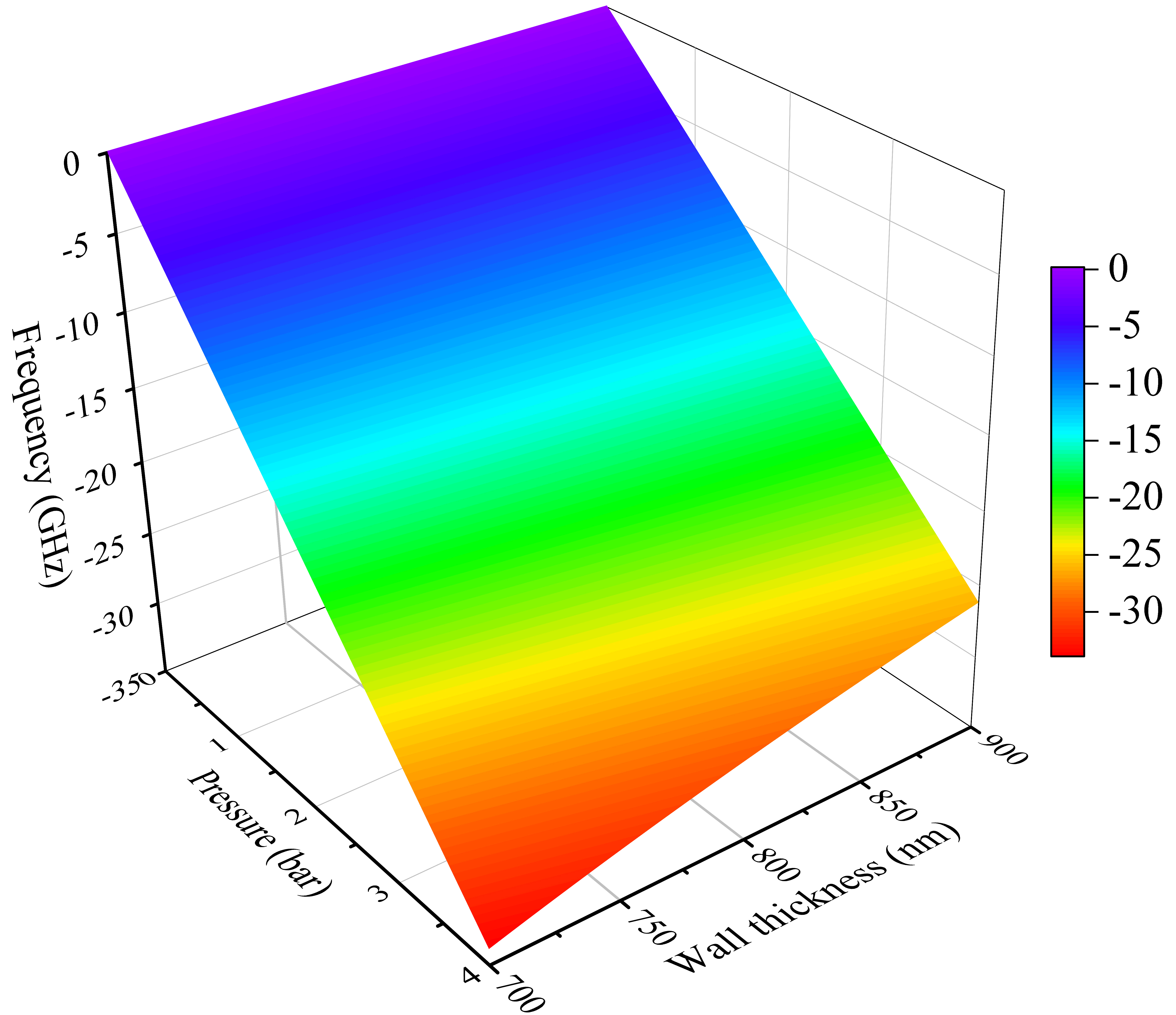}
  \caption{Simulated resonance frequency shift of a passive tellurite glass as a function of air pressure and wall thickness. The color bar represents the frequency shift in units of GHz. 
  \label{fig:Simulation}}
\end{figure}

In order to characterize the mechanical properties of the passive tellurite glass microbubbles, the elasticity equations described in \cite{Henze:11} were used to calculate the frequency shift of the resonances as a function of pressure. The required material parameters used were: shear modulus $G=27.5$ Gpa, bulk modulus $K=40$ Gpa, refractive index $n=2$, Young's modulus $E=67.2$ Gpa, elasto-optical constants $C_1=-1.8\times10^{-12}$ m$^{2}$/N and $C_2=-2\times10^{-12}$ m$^{2}$/N \cite{ELMALLAWANY199893,Weber2003HandbookOO}. Wall thicknesses from 700 to 900 nm were used in the calculation and the results are shown in Figure \ref{fig:Simulation}. Note that the theoretical value of air pressure sensitivity was between $-6.4$ to $-8.3$ GHz/bar for the wall thicknesses used. Compared with  silica, tellurite has higher shear and bulk moduli, but lower elasto-optical constants.  As a comparison, the pressure sensitivity of a silica microbubble with the same diameter and wall thicknesses was calculated to be $-8.7$ to $-11.2$ GHz/bar.

\section{Active Yb$^{3+}$-Er$^{3+}$ co-doped tellurite glass microbubble}
\begin{figure} [ht]
\centering
      \includegraphics[width=0.65\textwidth]{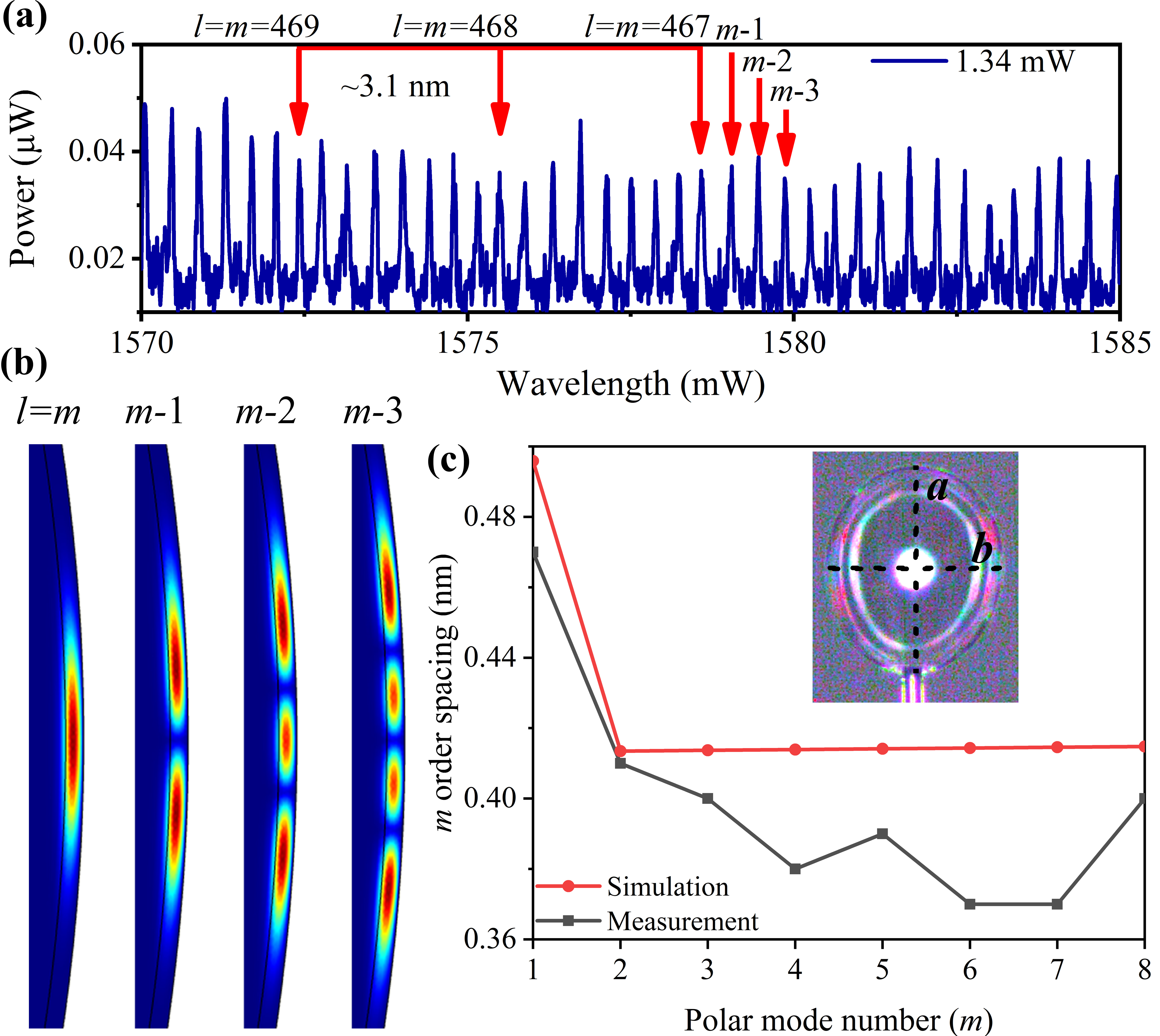}
  \caption{(a) The measured output spectrum when the pump power is below threshold; the red arrows indicate that the FSR is about 3.1~nm, \textit{l} and \textit{m} are the polar and azimuthal mode numbers in the microbubble. (b) The electric field distribution of different polar modes, when the mode number $l=467$. (c) The simulation and measurement of $m$ order spacing as function of polar mode number.  The inset is a microscope image of the microbubble, where \textit{a} and \textit{b} are the major and minor axes.
  \label{fig:pump}}
\end{figure}

Yb$^{3+}$-Er$^{3+}$ co-doped tellurite microbubbles (with diameters around 130~$\mu$m) were pumped using a 980~nm diode laser. When the pump laser output power was increased from 0 mW to 1.34~mW , a fluorescence spectrum was observed on the OSA. 
A free spectral range (FSR) of 3.1~nm was fitted to the fluorescence spectrum from a theoretical calculation result of 3.02~nm \cite{Sumetsky:10}. 
For this particular microbubble, we observe 8 modes in a single FSR, see Figure \ref{fig:pump}(a). The main reason for this is that the shape of the fabricated tellurite microbubble is not perfectly spherical, hence some polar modes are excited within.

\begin{figure} [ht]
\centering
      \includegraphics[width=0.9\textwidth]{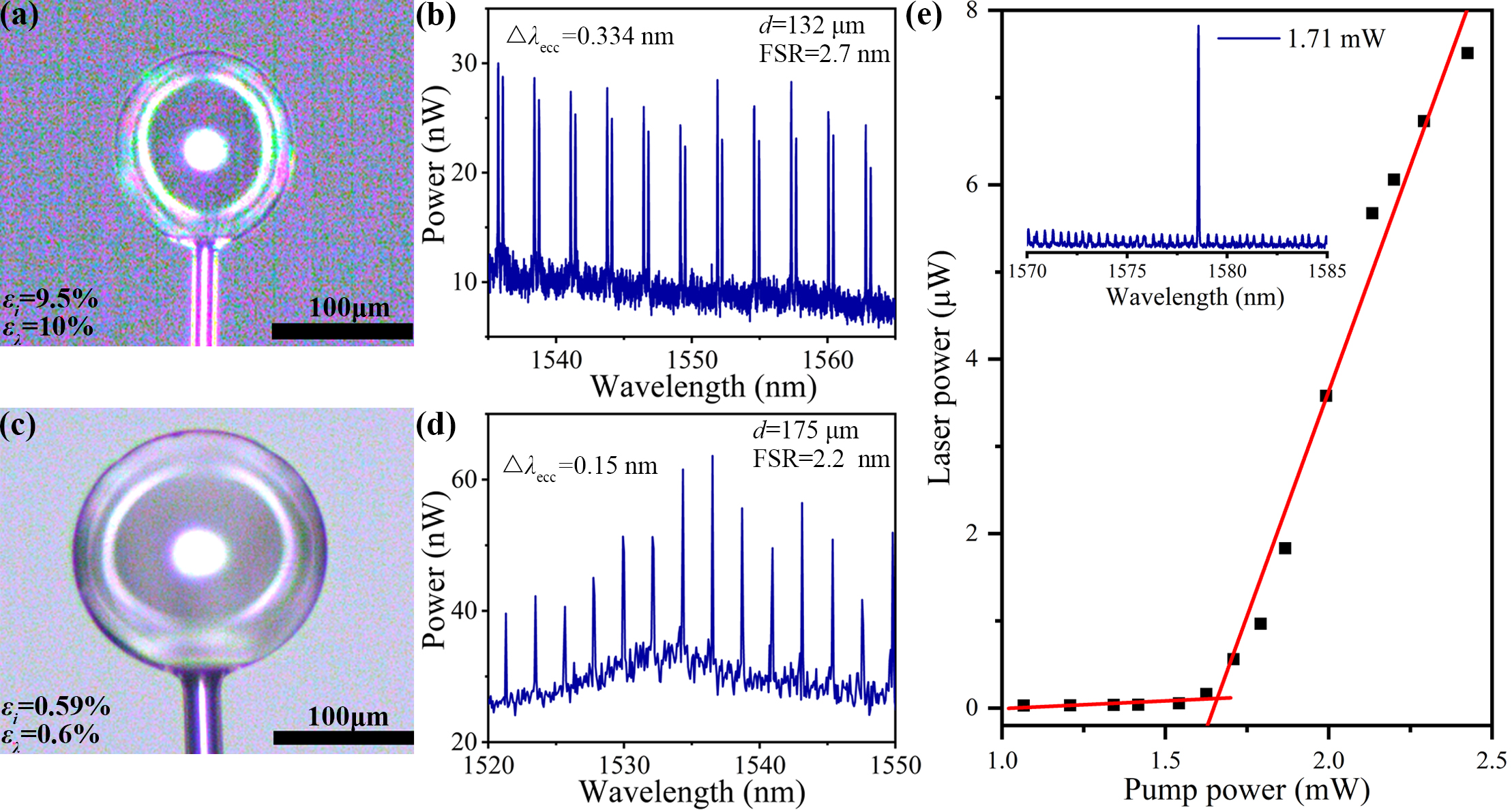}
  \caption{(a) and (c) Microscope images of the microbubbles under test. $\varepsilon_{i}$ and $\varepsilon_{\lambda}$ are the measured and calculated eccentricity of the microbubbles. Fluorescence spectrum of the Yb$^{3+}$-Er$^{3+}$ doped tellurite glass microbubble with only two modes (b) or single mode (d) emission in an FSR, corresponding to the bubbles in (a) and (c), respectively. \textit{d} is the diameter of the microbubble. $\Delta \lambda_{\rm ecc}$ is the wavelength spacing between the polar modes. (e) Laser power at 1578.56~nm as a function of pump power. The red line represents a linear fit to the experimental data, with a lasing threshold of 1.66~mW. The inset is the output laser spectrum when the pump power is 1.71 mW.
    \label{fig:fluor}}
\end{figure}

The mode number at 1578.56~nm was calculated to be $l=m=467$.  The number of field maxima  in the polar direction is given by $l-m+1 = 1,2,3...$ The first three of these polar modes are also highlighted in Figure \ref{fig:pump}(a) and have a measured mode spacing of 0.47 nm, 0.41 nm and 0.4 nm, respectively. The polar mode spacing is in close agreement with the calculated mode spacing determined from a numerical FEM (COMSOL) model, see Figure \ref{fig:pump}(b) and (c).    An image of the microbubble is given in the inset of Figure \ref{fig:pump}(c). If we define \textit{a} as the major axis and \textit{b} as the minor axis, the eccentricity, $\varepsilon_i$, of the microbubble can be calculated from \cite{Onchip2017}
\begin{equation}
\textit{$\varepsilon_{i}$}=\frac{\textit{a}-\textit{b}}{\textit{a}}. 
\end{equation}
Additionally, the eccentricity, $\varepsilon_{\lambda}$, determined from the mode spacing can be calculated from \cite{Onchip2017}

\begin{equation}
\Delta f_{\rm ecc}=|f_{ml}-f_{m+1l}|\approx f_{ml}\cdot\textit{$\varepsilon_{\lambda}$}\frac{\textit{|m|}-{1/2}}{\textit{l}},
\end{equation}

\noindent where $f_{ml}$ is the frequency of the $ml$ mode.  
The $\varepsilon_{i}$ and $\varepsilon_{\lambda}$ of the bubble were measured and calculated as 0.14 and 0.12, respectively, and are in reasonable agreement. Even though the walls of these microbubbles appear to be of uniform thickness, it is not surprising that the bubble shape can deform to a degree that lifts the mode degeneracy. Differently shaped, doped microbubbles were also tested by pumping with the 980~nm light. The resulting fluorescence spectra and images are shown in Figure \ref{fig:fluor}(a)-(d). As the eccentricity decreases, the number of higher order modes decreases, as expected. When the eccentricity is 10$\%$ only one higher order mode exists within a single FSR. As the eccentricity drops, the higher order mode spacing decreases and below 1$\%$ the mode degeneracy is nearly recovered resulting in a single mode spectrum.

Most compound glasses have larger refractive indices than silica. For example, when a tapered fiber is used to pump a rare-earth doped compound microsphere with a high refractive index, many higher order WGMs are excited \cite{Yu2018}. This could be attributed to the fact that the gain of many of the modes is greater than the loss, resulting in a lower energy conversion efficiency and the resonator being more prone to output mode hopping.  As the pump power was increased further to 1.63~mW, laser emission at a wavelength of 1578.56~nm was detected. The results are shown in Figure \ref{fig:fluor}(e), where the relationship between the 980~nm pump power and the detected power of a single WGM lasing mode is plotted. The lasing threshold is about 1.66~mW and single-mode lasing output was observed throughout the entire measurement cycle. 

\begin{figure} [ht]
\centering
      \includegraphics[width=0.8\textwidth]{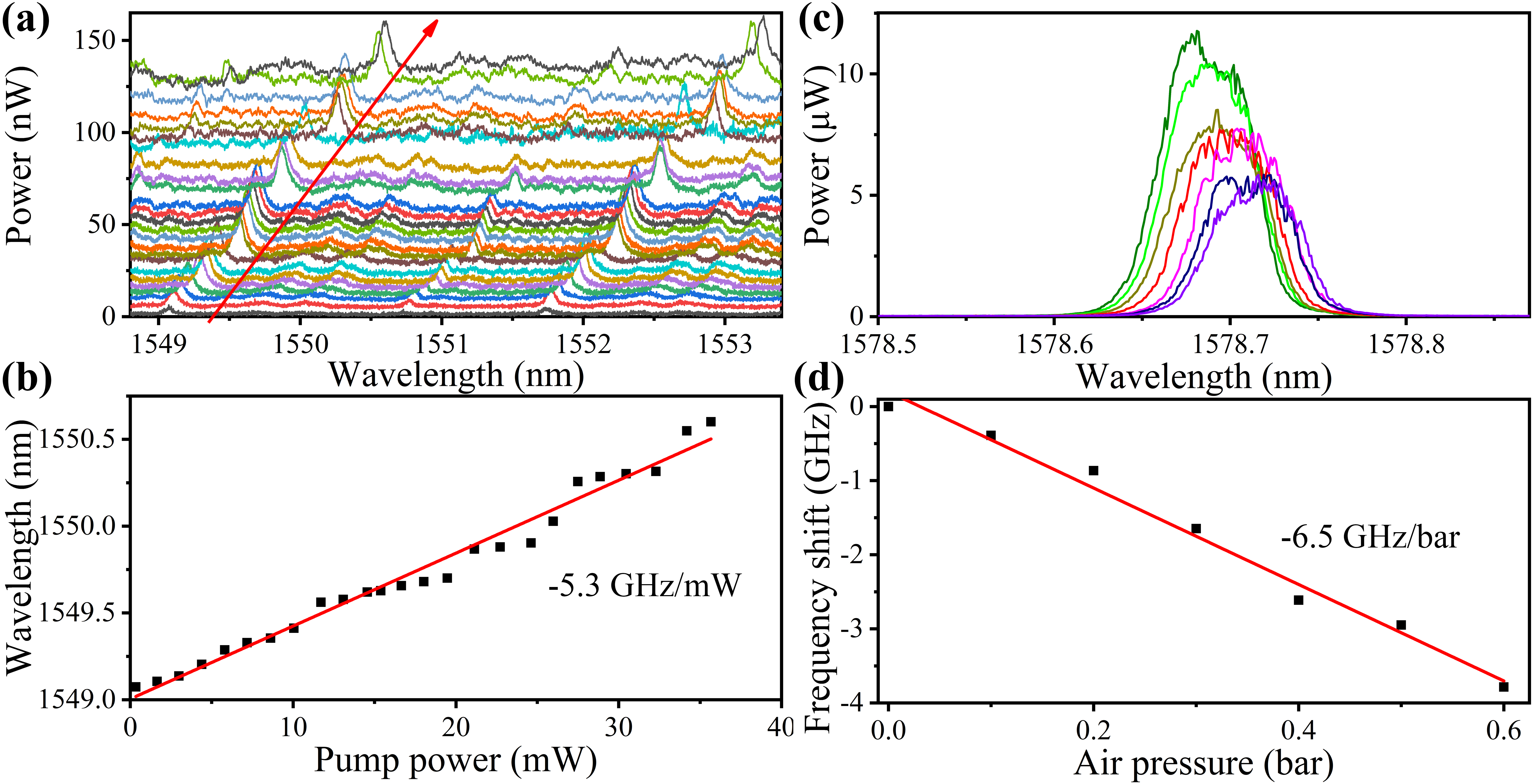}
  \caption{(a-b) The measured fluorescence emission spectrum obtained by increasing the pump power from 0 to 35.6~mW. The maximum wavelength tuning range is about 1.53~nm and the red arrow indicates the direction of wavelength tuning. (c) Lasing wavelength of the active tellurite glass microbubble at different pressures. (d) Frequency shift as a function of the internal air pressure. Red lines are linear fits.
  \label{fig:tuning}}
\end{figure}

Wavelength tuning of the Yb$^{3+}$-Er$^{3+}$ doped tellurite glass microbubble with a diameter of 130~$\mu$m was also investigated by varying the temperature and the internal air pressure. The 980~nm diode laser was used as pump  to obtain a fluorescence spectrum and the pump power was increased from 0 to 35.6~mW. The resulting spectra are shown in Figure \ref{fig:tuning}(a). It should be noted that when the power launched into the tapered fiber was increased to 40~mW, a large loss was induced because the tellurite glass microbubble melted and fused to the tapered fiber \cite{4451197}. The result of the tuning is shown in Figure \ref{fig:tuning}(b) and a maximum wavelength shift, that is tuning range, of 1.53~nm was obtained. Some jumps in wavelength tuning range were observed due to the thermal effects around pump/cavity resonances \cite{Carmon:s, Ward2010}. Although the tuning is in general nonlinear, the overall tuning rate is -5.3 GHz/mW.   The frequency shift of the WGM laser modes at different air pressures from 0 to 0.6~bar was also investigated and the results are shown in Figure \ref{fig:tuning}(c) and (d). A tuning sensitivity of $-6.5$~GHz/bar was determined following a linear fitting.  The accuracy is limited by the spectral resolution of the OSA.
  
\section*{Conclusion}
In this work, we report on a method to fabricate  microbubbles from a soft glass, namely tellurite, using a CO${_2}$ laser. The method involves first melting the glass capillary to form a sphere and then blowing out the sphere to make a bubble. The fabricated bubbles have diameters and wall thicknesses of approximately 150~$\mu$m and 670~nm, respectively. Notably, the bubbles have quite a uniform wall thickness and spherical shape which can result in a low number of higher order modes. The measured eccentricity can approach that previously observed in microspheres. The microbubbles were made from both passive and Yb:Er doped glass. The tellurite glass microbubbles were investigated experimentally and characterized in terms of their $Q$-factors, mode spectra, tuning rates, eccentricity, and laser output, and these results were compared against silica whispering gallery resonators where possible. In the case of passive microbubbles, a high \textit{Q}-factor of 2.3$\times$10$^6$ was achieved. A broadband 980~nm diode pump laser was used to change the temperature inside the microbubble and a large temperature sensitivity of 4.9~GHz/K was obtained. In addition, the air pressure sensitivity of 7.1~GHz/bar was  measured by applying different internal air pressures. Even though the tellurite glass is softer than silica, the expected increase in pressure tuning is negated by the lower elasto-optic coefficient.

For the active microbubbles, a maximum wavelength tuning range of 1.53~nm was observed by increasing the intensity of the pump light. Separately, the sensitivity of the output laser frequency to air pressure was determined as 6.5~GHz/bar which is similar to the undoped microbubble aerostatic tuning rate. Additionally, the devices fabricated in this investigation can have very low eccentricity and so fewer modes in a single FSR, resulting in a higher laser conversion efficiency. The research reported in this article has potential impact for many applications, including low threshold, high conversion efficiency and tunable microcavity laser sources operating in the near and mid-infrared range, integrated active photonic devices, and laser sensing using microbubble resonators based on compound glass. 

\section*{Acknowledgments}
The authors acknowledge the Engineering Support Section of OIST Graduate University. 

\section*{Funding}
This work was supported by the National Key Program of the Natural Science Foundation of China (NSFC 61935006), National Natural Science Foundation of China (NSFC 61905048, 61805074), the Fundamental Research Funds for the Central Universities (HEUCFG201841, GK2250260018, 3072019CFQ2503, 3072019CF2504, 3072019CF2506, 3072020CFJ2507, 3072020CFQ2501, 3072020CFQ2502, 3072020CFQ2503, 3072020CFQ2504), the 111 project (B13015) to Harbin Engineering University, Heilongjiang Provincial Natural Science Foundation of China ( LH2019F034). Heilongjiang Touyan Innovation Team Program, Harbin Engineering University Scholarship Fund and OIST Graduate University. 
\section*{Disclosures}
The authors declare no conflicts of interest.
\end{document}